\documentclass[12pt]{article}
\begin{document}

\title{{\bf Preliminary Inconclusive Hint of Evidence
Against Optimal Fine Tuning of the Cosmological Constant
for Maximizing the Fraction of Baryons Becoming Life}
\thanks{Alberta-Thy-02-11, arXiv:1101.2444[hep-th]}}

\author{
Don N. Page
\thanks{Internet address:
profdonpage@gmail.com}
\\
Theoretical Physics Institute\\
Department of Physics, University of Alberta\\
Room 238 CEB, 11322 -- 89 Avenue\\
Edmonton, Alberta, Canada T6G 2G7
}

\date{2011 January 28}

\maketitle
\large
\begin{abstract}
\baselineskip 25 pt

The effective coupling `constants' of physics, especially the
cosmological constant, are observed to have highly biophilic values.  If
this is not a hugely improbable accident, or a consequence of some
mysterious logical necessity or of some simple principle of physics, it
might be explained as a consequence either of an observership selection
principle within a multiverse of many sets of effective coupling
constants, or else of some biophilic principle that fine tunes the
constants of physics to optimize life.  Here a very preliminary
inconclusive hint of evidence is presented against the hypothesis of
optimal fine tuning of the cosmological constant by a biophilic
principle that would maximize the fraction of baryons that form living
organisms or observers.

\end{abstract}

\normalsize

\baselineskip 17 pt

\newpage

\section{Introduction}

It has long been recognized that many of the apparent constants of
physics are observed to take values that are much more biophilic (in the
sense of being conducive to life and observership) than values
significantly different are believed to be \cite{Carter, Carr-Rees,
Davies, Barrow-Tipler, Rees1, TARW, Carr}.  For example, the cosmological
constant (or dark energy density) that quantifies the gravitational
repulsion of empty space is roughly 122 orders of magnitude smaller than
the Planck value, but if it were just a few orders of magnitude larger
than its tiny positive observed value \cite{Perl, Riess}, with the other
constants of physics kept the same, life as we know it would appear to
be very difficult.  

A partial explanation for this apparent fine tuning is the anthropic
principle \cite{Carter, Carr-Rees, Davies, Barrow-Tipler, Rees1, TARW,
Carr}, that as observers we can observe only conditions (including the
constants of physics) that permit our existence.  However, it has been
controversial what the deeper implications of this are.

One view is it is purely an accident or coincidence that the constants
of physics have biophilic values, and that there is no deeper
explanation.  However, the fact that the cosmological constant is
roughly 122 orders of magnitude smaller than the apparently simplest
natural nonzero value for it (the Planck value) cries out for an
explanation beyond pure coincidence, since the probability of such a
remarkable coincidence from a random selection of the cosmological
constant with a measure uniformly distributed over a range roughly the
Planck value is extremely low, much less than the probability of having
a monkey randomly type on a simple typewriter in one go,
\begin{verbatim}
The cosmological constant is 10^(-122) in Planck units.
\end{verbatim}

A second view is that there are simple principles of physics, not
specifically connected with life, that uniquely determine the constants
to have the values they are observed to have.  This view was commonly
thought to be the case for the cosmological constant when it was
believed to be zero (though no compelling simple principle was ever
found that clearly implied that it would be zero).  However, now that
the cosmological constant has been found to have a very small positive
value, it is hard to see how a simple principle of physics, independent
of life, would uniquely fix that tiny nonzero value.

Perhaps a more common view among physicists today is the idea that there
is a multiverse with a wide range of values for the constants of
physics, and by the selection principle of observership (the weak
anthropic principle), we find ourselves in the part of the multiverse
where life is possible and/or relatively common (at least compared to
other parts of the multiverse) \cite{Carr}.  However, there is still
considerable controversy over whether such a multiverse that would be
necessary for this explanation really exists.

A fourth view is that there is some principle the fixes the constants of
physics as they are in order that life would occur in the universe,
perhaps in some maximal way.  One version of this is the idea that ``A
theist would not be surprised if God had optimized the universe for
life, not merely made life possible'' \cite{Polkinghorne-Beale}. The
scientific form of this hypothesis would be that there is some biophilic
property that is maximized by the constants of physics, so that the
constants of physics are optimally fine tuned for life in some specific
way.

The first view outlined above, pure coincidence, is hard to prove or
disprove, though it would seem implausible and would be hard to maintain
if any of the other three explanations were found to work well.  The
second idea, unique determination by simple principles of physics, could
in principle attain strong evidence for it if such principles were found
and were convincingly shown to lead to the observed constants of
physics.  It would be hard to disprove, since one could always believe
that there were such principles that simply had not yet been discovered,
but the more that scientists fail to find such principles, the less
attractive this option will seem, particularly if some of the other
options appear to work.  The third view, of observer selection within a
multiverse, is hard to prove or disprove directly, since it appears very
difficult to obtain direct information about other possible parts of a
multiverse.  However, if a simple theory were developed that gives good
statistical explanations for what we do observe and that also predicts a
multiverse that we cannot directly observe, such a theory could become
highly convincing (analogous to the prediction by general relativity of
very high curvature in black-hole interior regions that cannot be
directly observed).  The fourth view, a biophilic principle of fine
tuning for life, is also hard to prove or disprove, but if it is cast
into the prediction that a certain calculable quantity is maximized,
then it could also be scientifically testable.

In this paper I shall examine a particular variant of the fourth view,
that the fraction of baryons that develops into living organisms is
maximized by the observed constants of physics.  This hypothesis is in
principle falsifiable, and I shall argue that considerations of
hypothetical variations of the cosmological constant give a very
preliminary inconclusive hint that it may become falsified, since the
fraction of baryons that condense into galaxies that in turn form living
organisms would be higher if the cosmological constant were lower.  This
result thus gives a preliminary suggestion that there might eventually
be evidence against optimal fine tuning for life (or at least for
maximizing the fraction of baryons that become living organisms) by such
a biophilic principle.  

However, email comments by Robert Mann \cite{Mann}, Michael Salem
\cite{Salem}, and Martin Rees \cite{Rees} have shown me that it is not
at all clear that the very small increase in the fraction of baryons
that would condense into galaxies if the cosmological constant were zero
instead of its tiny observed positive value would also lead to an
increase in the fraction of baryons that would go into life, since
conceivably the small differences in the galaxies produced by lowering
the cosmological constant to zero might also affect the fraction of
baryons within galaxies that become life in a way that would
overcompensate for the higher fraction of baryons condensing into
galaxies.  

One should also recognize, as Nicholas Beale \cite{Beale} has emphasized
to me, that the total number of baryons might depend on the cosmological
constant in such as way as to make the total amount of life with a
smaller cosmological constant lower, even if the increase in the
fraction of baryons condensing into galaxies really does lead to an
increase in the fraction of baryons incorporated into living organisms. 
Although I shall discuss the latter issue below, for the most part in
this paper I am essentially ignoring it and merely considering the
hypothesis that the cosmological constant is optimally fine tuned to
maximize the fraction of baryons becoming living organisms.

A further caveat stressed by Robin Collins \cite{Collins} and Beale
\cite{Beale} is that it is by no means clear that it would be the total
amount of life that would be optimized, but perhaps intelligent life
\cite{Gleiser}, ``life that develops technology and science''
\cite{Collins}, ``life that is capable of real love'' \cite{Beale},
``life that could discover the laws of physics''
\cite{Beale,Bealepaper}, etc.  So even if one could show that there is
only a single universe with constants of physics that could be fine
tuned for life but were not optimally fine tuned to maximize life, this
by itself would not be evidence against the hypothesis that the
constants are optimally fine tuned to maximize something having to do
with life.  However, to formulate a conjecture that in principle can be
tested scientifically, it seems one needs to make it much sharper than
can be justified by any such arguments, and in this paper it is only the
very specific hypothesis that the constants of physics are optimally
fine tuned to maximize the fraction of baryons forming living organisms
that is considered.  Here I am also making the very strong assumption
that the cosmological constant can indeed be varied entirely
independently of all the other constants of physics that may be
responsible for all microphysics, including inflation, flatness, density
fluctuations, baryosynthesis, nucleosynthesis, relative particle
numbers, particle masses, coupling constants, stellar evolution,
chemistry, geophysics, biology, Darwinian evolution by natural
selection, life, complex life, intelligence, observers, observation,
consciousness, etc.

\baselineskip 16.5 pt

\section{The positive cosmological constant as a very preliminary
suggestive hint of evidence against fine tuning for maximizing the
fraction of baryons in living organisms or observers}

Martel, Shapiro, and Weinberg \cite{MSW}, following upon previous ideas
of Weinberg \cite{Wein}, have shown that the fraction of baryons that
condense gravitationally into structures large enough to form living
observers is a very sensitive function of the cosmological constant
$\Lambda$ that decreases rapidly if $\Lambda$ is much larger than the
observed value $\Lambda_O$ (which is about $3.5\times 10^{-122}$ in
Planck units, $\hbar = c = G = 1$). Therefore, if indeed there is a
multiverse with a wide range of values of $\Lambda$ that are fairly
uniformly distributed near $\Lambda = 0$, the third view, observer
selection within a multiverse, would be a good explanation for the
observed value $\Lambda_O$.

However, here we are examining the alternative hypothesis from the
fourth view, that a biophilic principle optimally fine tunes $\Lambda$
to the value that maximizes the fraction of baryons that develop into
life.  Martel, Shapiro, and Weinberg \cite{MSW} found that not only does
the fraction of baryons condensing into galaxies drop steeply with
$\Lambda$ if it is much larger than $\Lambda_O$, but also that it is a
decreasing function of $\Lambda$ for all positive values. The simple
reason, which does not depend on any of the details of the particular
nonlinear model Martel, Shapiro, and Weinberg used, is that a positive
cosmological constant gives a repulsion between distant particles that
reduces the ordinary gravitational attraction and leads to less
gravitational condensation of matter.  Therefore, any positive
cosmological constant decreases the fraction of baryons that condense to
form galaxies and other structures that eventually may form living
substructures.

As a consequence, if the fraction of baryons that form living organisms
out of the baryons that condense into galaxies is independent of the
fraction of baryons that condense into galaxies out of all baryons, no
positive value of the cosmological constant (such as the observed value
$\Lambda_O$) can maximize the fraction of baryons in life.  Therefore,
under the assumption of the previous sentence (that a fixed fraction of
the baryons that condense into galaxies become living organisms), the
observed positive value of the cosmological constant is evidence against
the specific hypothesis of optimally fine tuning for life by a biophilic
principle that would maximize the fraction of baryons that form living
organisms or observers.

However, I have been informed by Robert Mann \cite{Mann}, Michael Salem
\cite{Salem}, and Martin Rees \cite{Rees} that it is plausible that the
fraction of baryons within galaxies that form life depends on the
cosmological constant (perhaps through the fraction of baryons that
condense into galaxies) in such a way that it might reverse what would
otherwise be the increase in the fraction of all baryons forming life
with a decrease of the cosmological constant below its observed value. 
For example, Mann \cite{Mann} raised the possibility that if a larger
fraction of baryons collapses, then perhaps statistically the
gravitational bound states that form are more biophobic, with the
stronger gravity of these structures making it less likely for life to
survive.  Salem \cite{Salem} agreed that the collapse fraction of
baryons is greater when the cosmological constant has a smaller
magnitude but pointed out, ``It seems plausible to me that the chance of
a disruptive astrophysical event is larger if the rate of baryon
accretion is larger.''  Similar but greater effects from a negative
cosmological constant were noted in his earlier paper \cite{SGSV}.  Rees
\cite{Rees} emphasized, ``there are other astrophysical complexities:
e.g.\ how much gas is retained in galaxies, despite the injection of
energy from stellar winds and supernovae, which is then able to form
later-generation stars containing heavy elements,'' and concluded that
the small effect of the observed value of the cosmological constant on
the fraction of baryons that condense into galaxies ``is likely to be
swamped by much bigger factors that we can't quantify,'' so that ``it's
not obvious that the trend is monotonic.''

As Rees \cite{Rees} later noted, ``I agree with you that if lambda is
below the critical value that suppresses galactic scale structure
completely (and we know it is several times below that), nothing should
depend very much on its value (unless the probability of life spreading
through the galaxy happened to be exceedingly sensitive to some feature
of galaxies that might be a function of lambda).''  Therefore, the issue
comes down to comparing several small effects.  Originally I just
considered the effect of the cosmological constant on the fraction of
baryons that condense into galaxies, say $F_G$, and assumed that the
fraction of baryons in galaxies that become life, say $F_{LG}$, is
rather insensitive to the cosmological constant, so that the fraction of
baryons that become life, say $F_L$, is proportional to $F_G$ with a
constant proportionality factor $F_{LG}$ and hence decreases
monotonically with the cosmological constant for positive values less
than the observed value.  That is, I was assuming that $F_L = F_G
F_{LG}$ with $F_{LG}$ rather independent of the fraction $F_G$ that
increases as $\Lambda$ is reduced below its observed value $\Lambda_0$
(also assuming, as I shall continue to do, that life develops only from
baryons that condense into galaxies).

Let $y=\Omega_\Lambda$, the ratio of the energy density of the
cosmological constant to the critical density at the present observed
value of the Hubble constant $H_0$ (which is assumed to be held fixed
under a hypothetical variation in the cosmological constant, so that the
cosmological constant is a fixed constant of proportionality multiplied
by $y$, which I shall thus use as a measure of the cosmological
constant).  In what follows, the derivatives with respect to $y$ are
partial derivatives keeping all the other constants of physics fixed,
but since in this paper I am only considering varying the cosmological
constant with the other constants held fixed, they are also ordinary
derivatives under this assumed restriction.  Then
$\partial(\ln{F_L})/\partial y = \partial(\ln{F_G})/\partial y +
\partial(\ln{F_{LG}})/\partial y$.  Simple reasoning that is confirmed
by Martel, Shapiro, and Weinberg \cite{MSW} (but which does not depend
on their detailed assumptions) shows that $\partial(\ln{F_G})/\partial y
< 0$ for positive $y$ at least, so if one neglects
$\partial(\ln{F_{LG}})/\partial y$, then $\partial(\ln{F_L})/\partial y
< 0$ for positive $y$, implying that the maximum of $F_L(y)$ cannot be
at the positive observed value of $y$.  However, Mann \cite{Mann}, Salem
\cite{Salem}, and Rees \cite{Rees} noted effects that might make
$\partial(\ln{F_{LG}})/\partial y > 0$.  The open question is whether it
could be that $\partial(\ln{F_{LG}})/\partial y >
|\partial(\ln{F_G})/\partial y| = -\partial(\ln{F_G})/\partial y$, in
which case $\partial(\ln{F_L})/\partial y > 0$ for small $y$.  Then
since presumably $\partial(\ln{F_L})/\partial y < 0$ for sufficiently
large $y$ (where $F_G(y)$ does become strongly suppressed by the
repulsive effects of a large cosmological constant in a way that seems
difficult to be compensated by any plausible increase in $F_{LG}(y)$),
the maximum of $F_L(y)$ would be at positive $y$ and in principle might
possibly be at the observed value.

It is still not clear to me even what the sign of
$\partial(\ln{F_{LG}})/\partial y$ is for very small positive values of
the cosmological constant.  One might na\"{\i}vely guess that in a
galaxy there would be a biophilic region where the metallicity is
suitable for life, and where the density of stars is not so great that
there are significant perturbations to the orbits of planets (so they
can have rather stable temperatures).  Presumably the centers of
galaxies would not be very good for life from the latter effect, so
perhaps it is not surprising that we do not find ourselves there.  But
then perhaps one could use as a surrogate of life, not merely the
fraction $F_G$ of baryons that condense into galaxies, but the fraction
of baryons, say $F_S$, that condense into stars that are sufficiently
similar to our sun in luminosity and metallicity and in not having too
large a surrounding density of other stars.  (Beale
\cite{Beale,Bealeblog} has alternatively suggested Habitable Earth-Like
Planets or HELPs, though to maximize their expected number rather than
to maximize the fraction of baryons that condense into them.)  Even if
galaxies are larger from the effects of more baryons condensing when $y$
is smaller, it is not obvious to me that this fraction $F_S$ of baryons
that form suitable stars would be reduced, since even though the centers
of such galaxies might be worse for life, conceivably the biophilic
regions far away for the centers might have more suitable stars (and
solar systems) for life if the galaxy as a whole has more baryons.  But
it is difficult to see how to estimate this effect, or other effects
such as the amount of gas retained by the galaxies.  Therefore, at
present it seems one should just say that there are effects that might
indeed make $\partial(\ln{F_{LG}})/\partial y >
|\partial(\ln{F_G})/\partial y|$ for small $y$, so that the maximum of
$F_L(y)$, the fraction of baryons that form life, could be at positive
$y$ and conceivably even at the observed value.

\section{Discussion and conclusions}

If we could become convinced that for positive cosmological constant the
derivative with respect to the cosmological constant (with all the other
constants of physics held fixed) of the logarithm of the fraction
$F_{LG}$ of baryons in galaxies that become life is smaller than the
absolute value of the (negative) derivative of the logarithm of the
fraction $F_G$ of all baryons that condense into galaxies, i.e., that
$\partial(\ln{F_{LG}})/\partial y < |\partial(\ln{F_G})/\partial y| =
-\partial(\ln{F_G})/\partial y$, then we would have evidence from the
positive observed value of the cosmological constant that this constant
of physics has not been optimally tuned to maximize the fraction $F_L$
of baryons going into life.  However, the fact that this inequality is
uncertain implies that the present argument merely gives a preliminary
inconclusive hint of evidence against optimal fine tuning of the
cosmological constant for maximizing the fraction of baryons becoming
life.  Unless the cosmological constant is in fact optimally fine tuned
in this way, it would seem that in principle one should be able to find
that out by calculating $\partial(\ln{F_L})/\partial y =
\partial(\ln{F_G})/\partial y + \partial(\ln{F_{LG}})/\partial y$ at the
observed value of $y=\Omega_\Lambda$ and showing that it is nonzero. 
Therefore, the hypothesis of optimal fine tuning is in principle
falsifiable, but unfortunately it appears to be premature at present to
be able to perform the proper calculations or tests.

Another potential objection to the preliminary inconclusive hint of
evidence against optimal fine tuning given here is that conceivably
there are simple principles of physics that give relationships between
the cosmological constant and other constants of physics that also
affect life, so that if these principles are upheld (rather than
allowing the cosmological constant to be varied independently of all the
other constants of physics in the fine tuning as was assumed above),
even if it can be shown that $\partial(\ln{F_L})/\partial y \neq 0$ at
the observed value of $y$ with all of the other constants of physics
held fixed, it might still possibly turn out that the cosmological
constant and the other constants of physics in fact do maximize the
fraction of baryons that become living organisms, subject to the
constraints of the principles of physics that give the putative
relationships between the constants.  However, if these principles of
physics allow variation of the constants, it is hard to see why they
would not allow the cosmological constant to be varied rather
independently from the other constants upon which life depends.  For
example, in the string landscape, it appears that the cosmological
constant can vary rather independently of the other constants of
physics.

A bigger objection, further emphasized by Beale \cite{Beale}, is it
might seem unreasonable to maximize the fraction of baryons becoming
life rather than just some measure of the totality of life itself.  If
the total number of baryons in the universe, say $B$, were some function
of $\Lambda$ or $y$ (with the other constants of physics held fixed at
their observed values), then it would seem more reasonable to maximize
something like the total number of baryons that form life (or perhaps
life of some particular kind), say $L = B F_L$, rather than the fraction
$F_L$.  However, we do not know any such dependence of the total number
of baryons on $\Lambda$, so in order to do their calculations, Martel,
Shapiro, and Weinberg \cite{MSW} left out that unknown dependence and
just considered the fraction of baryons condensing into structures, and
similarly I can also do little other than to leave it out. 
Nevertheless, that uncertainty is certainly cause for worry
\cite{Weinberg-in-Carr}. The situation seems to be even more ambiguous
by the fact that the simplest estimates for the total number of baryons
tend to be infinite.  This leads us to the whole measure problem in
cosmology \cite{Linde86, LinMez, Vil95a, LLM, Guth00, SGSV, PageJCAP08,
BFYb, DGLNSV, SPV, BFLR, Agnesi}, which I admit has no universally
accepted solution despite my own present favorite partial solution
\cite{Agnesi}.

However, one can formulate the following crude rebuttal argument to the
argument that it is plausible that $B$ depends on $\Lambda$ or $y$ in
just such a way that $L$ is maximized at the observed value of $y$ even
if the maximum of $F_L(y)$ is not near the observed value of $y$:  Since
we have very little idea about $B(y)$, it seems that our uncertainty of
$\ln{|\partial(\ln{B})/\partial y|}$ is large.  At the negative end, if
the only dependence of $B$ on $y$ is from the very tiny relative change
of the expansion rate of the universe during baryosynthesis, one would
have $|\partial(\ln{B})/\partial y| \ll |\partial(\ln{F_L})/\partial
y|$, so then we could ignore the variation of $B(y)$.  At the positive
end, one could have $|\partial(\ln{B})/\partial y| >>
|\partial(\ln{F_L})/\partial y|$ for small $|y|$, in which case the
maximum for $L(y)$ would presumably be for $|y|$ much larger than the
observed value, again contrary to the hypothesis that the observed value
maximizes $B(y)$.  Assuming that $|\partial(\ln{F_{LG}})/\partial y|$ is
not much larger than $|\partial(\ln{F_G})/\partial y|$ that is small in
comparison with unity, so that $|\partial(\ln{F_L})/\partial y| \sim
|\partial(\ln{F_G})/\partial y| \ll 1$, for $B(y)$ to be maximized at
the observed value would require that $\ln{|\partial(\ln{B})/\partial
y|} \sim \ln{|\partial(\ln{F_G})/\partial y|}$, which seems a priori
rather unlikely, given the large uncertainty in
$\ln{|\partial(\ln{B})/\partial y|}$.

It might be appropriate to note that although this paper has focused on
the scientifically testable question of whether the constants of physics
maximize a particular measure for life, it obviously also has
theological implications.  It could be taken as a preliminary
inconclusive hint of negative evidence for theists who expect God to
fine tune the constants of physics optimally for the fraction of baryons
going into life \cite{Polkinghorne-Beale} (though even these authors do
not expect such an optimal tuning merely for the fraction of baryons
going into life, or even for just the total amount of life without
regard to its quality, so I have taken their writings as motivation for
a much stronger hypothesis in order that it might be testable). 
However, for other theists, such as myself, even if it could be shown
that the constants of physics are not optimally fine tuned to maximize
any reasonable measure of life in a single universe with just one set of
constants, such evidence may simply support the hypothesis that God
might prefer a multiverse with many sets of constants as the most
elegant way to create life and the other purposes He has for His
Creation \cite{Does-God}.

In conclusion, the fact that the observed cosmological constant is
positive may be taken to be a very preliminary inconclusive hint of
evidence against a biophilic optimal fine tuning of it to maximize the
fraction of baryons that develop into living organisms, since to
maximize that fraction, the simplest (but highly uncertain) assumption
would be that the fraction of baryons condensing into galaxies would
need to be maximized, and for that the cosmological constant would
instead need to be slightly negative.

I am grateful for many email responses to an earlier version of this
paper from (in chronological order of the first response from each
person) Andrei Linde, Robert Mann, Robert Collins, Nicholas Beale,
Michael Salem, Martin Rees, Hugo Martel, and Paul Shapiro.  I also
appreciate the hospitality of the Princeton Center for Theoretical
Science, where I had highly stimulating further discussions on this and
related subjects with Andreas Albrecht, Tom Banks, Adam Brown, Sean
Carroll, Ben Freivogel, Jaume Garriga, Alan Guth, Jean-Luc Lehners, Juan
Maldacena, Roger Penrose, Paul Steinhardt, Neil Turok, Alexander
Vilenkin, and Edward Witten.  This work was supported in part by the
Natural Sciences and Engineering Research Council of Canada.

%\newpage

\end{document}